\RequirePackage{ifpdf}
\ifpdf 
\documentclass[pdftex]{sigma}
\else
\documentclass{sigma}
\fi

\newcommand{\mib}[1]{\mbox{\boldmath $#1$}}

\def\bra{\langle}
\def\ket{\rangle}

\def\d{\delta}
\def\e{\epsilon}

\def\l{\lambda}
\def\m{\mu}

\def\r{\rho}
\def\s{\sigma}

\def\nn{\nonumber}
\def\sh{{\rm sinh}}

\def\ds{\displaystyle}
\def\ov{\overline}
\def\({\left(}
\def\){\right)}
\def\[{\left[}
\def\]{\right]}

\def\nonum{\nonumber}

\newcommand{\nc}{\newcommand}
\nc{\za}[1]{\zeta_a(#1)}
\nc{\zcor}[2]{\bra S_{#1}^z S_{#2}^z \ket}

\begin{document}

\allowdisplaybreaks

\renewcommand{\thefootnote}{$\star$}

\renewcommand{\PaperNumber}{004}

\FirstPageHeading

\ShortArticleName{Correlation Function and Simplif\/ied TBA Equations for XXZ Chain}

\ArticleName{Correlation Function and Simplif\/ied TBA Equations\\ for XXZ Chain\footnote{This paper is a
contribution to the Proceedings of the International Workshop ``Recent Advances in Quantum Integrable Systems''. The
full collection is available at
\href{http://www.emis.de/journals/SIGMA/RAQIS2010.html}{http://www.emis.de/journals/SIGMA/RAQIS2010.html}}}

\Author{Minoru TAKAHASHI}

\AuthorNameForHeading{M.~Takahashi}

\Address{Fachbereich C  Physik, Bergische Universit\"at Wuppertal, 42097 Wuppertal, Germany}
\Email{\href{mailto:mtaka@issp.u-tokyo.ac.jp}{mtaka@issp.u-tokyo.ac.jp}}

\ArticleDates{Received September 27, 2010, in f\/inal form December 27, 2010;  Published online January 08, 2011}

\Abstract{The calculation of the correlation functions of Bethe ansatz solvable models
is very dif\/f\/icult problem. Among these solvable models spin 1/2 XXX chain has
been investigated for a long time. Even for this model only the nearest
neighbor and the second neighbor correlations were known.
In 1990's  multiple integral formula for the general
correlations is derived. But the integration of this formula is also very dif\/f\/icult problem.
Recently these integrals are decomposed
to products of one dimensional integrals and at zero temperature, zero magnetic f\/ield and isotropic case,
correlation functions are expressed by $\log 2$ and Riemann's zeta functions with odd integer argument
$\zeta(3), \zeta(5), \zeta(7),\dots$.
We can calculate density sub-matrix of successive seven sites.
Entanglement entropy of seven sites is calculated.
These methods can be extended to XXZ chain up to $n=4$. Correlation functions
are expressed by the generalized zeta functions.
  Several years ago I derived new thermodynamic Bethe ansatz equation for XXZ chain.
This is quite dif\/ferent
with Yang--Yang type TBA equations and contains only one unknown function.
This equation is very useful to get the high temperature expansion.  In this paper we get the analytic
solution of this equation at $\Delta=0$.}


\Classification{16T25; 17B37; 82B23}

\section{Introduction}
We consider the spin 1/2 XXZ chain
\begin{gather}
{\cal H}=-J \sum_{l=1}^N S_l^xS_{l+1}^x+S_l^yS_{l+1}^y+\Delta \left(S_l^zS_{l+1}^z-\frac{1}{4}\right)-2h\sum_{l=1}^NS_l^z.
\label{XXZham}
\end{gather}
Among the solvable models this model has been investigated for a long time.

It was believed that the exact calculation of correlation function
is impossible except the nearest neighbor correlation function.  I derived the
second neighbor correlation function for $J<0$, $T=0$, $h=0$, $\Delta=1$ using
the Lieb--Wu solution of one-dimensional Hubbard model~\cite{Takahashi77}.
Details are  given in Appendix~\ref{sce}.
In 1990's  multiple integral formula were
proposed for $h=0$, $T=0$  and recently multiple integral formula was extended to
$h\ne 0$, $T\ne 0$.  But the factorization of these multiple integrals to the integrals of
lower dimension still remains a dif\/f\/icult problem.  I~explain the present situation
of factorization in Section~\ref{cf}.

Since Yang and Yang proposed the thermodynamic Bethe ansatz (TBA) equation for one dimensional bosons
\cite{Yang69},
the calculation of free energy becomes important for other solvable models.  Yang--Yang
type integral equations for \eqref{XXZham} were proposed in early 1970's~\cite{Taka71, Gaudin71, Taka72}.
In this theory inf\/inite number of unknown functions appeared for  $|\Delta|\geq 1$
and f\/inite number of unknowns appear for $\Delta=\cos \pi \nu$ with $\nu={\rm rational\ number}$.
The equations change by the value of $\Delta$ and not so convenient for numerical calculations.
But some important physical properties at low temperature were investigated using these equations.

Around 1990 the quantum transfer matrix method was applied to this model and numerical results coincide with those of Yang--Yang type equations~\cite{Koma89,Taka91,Klum93}.
In 2001 I proposed a new TBA equation which contains only one unknown function for \eqref{XXZham}~\cite{mtaka01d,mtaka01e}. This equation is very convenient to do the high-temperature expansions and we get one hundred-th order of high temperature expansion~\cite{shiro02}. The
traditional  cluster expansion method gives up to 22'nd. This method was also applied to the Perk--Schultz model~\cite{Tsuboi05}.

Unfortunately this equation does not numerically converge at
low temperature like $T/|J|<0.07$ because the integrand strongly oscillates. Some other numerical
method is necessary. I~showed that this equation gives the known exact results in Ising limit
($\Delta\to \infty$ and $J\Delta={\rm f\/inite}$)~\cite{mtaka01d}. In Section~\ref{sim} I give the analytic solution of
this equation for XY case ($\Delta=0$).

\section{Correlation functions of XXZ chain}\label{cf}

\subsection[Known exact results for $J<0$, $T=0$, $h=0$, $\Delta=1$]{Known exact results for $\boldsymbol{J<0}$, $\boldsymbol{T=0}$, $\boldsymbol{h=0}$, $\boldsymbol{\Delta=1}$}

\begin{enumerate}\itemsep=0pt
\item Nearest-neighbor correlator
\begin{gather}
  \langle S_j^z S_{j+1}^z \rangle = \frac{1}{12}-\frac{1}{3} \ln 2 = -0.1477157268 \dots \label{nn38}
\end{gather}
from the ground state energy per site by Hulth\'{e}n~\cite{Hulthen38}
 (1938).
\item Next nearest-neighbor correlator for XXX
\begin{equation}
\langle S_j^z S_{j+2}^z \rangle = \frac{1}{12}-\frac{4}{3} \ln 2 + \frac{3}{4} \zeta(3)
=0.06067976995 \dots, \label{nnn}
\end{equation}
 from the ground state energy of the half-f\/illed Hubbard model by Takahashi~\cite{Takahashi77} (1977).
\item The twisted four-body correlation function
\begin{gather}
  \langle
    (\mib{S}_{j  } \times \mib{S}_{j+1})\cdot
    (\mib{S}_{j+2} \times \mib{S}_{j+3})
  \rangle
  =  \frac{1}{2}\ln2 - \frac{3}{8}\zeta(3)=-0.104197748 \label{twist}
\end{gather}
from the third derivative of transfer matrix  by Muramoto and Takahashi~\cite{mura99} (1999).
\end{enumerate}

\subsection{Multiple-integral representations for density matrix}

For more general XXZ model with an anisotropy parameter ${\Delta}$

\begin{enumerate}\itemsep=0pt
\item
$ \Delta>1$, $T=0$, $h=0  $:  Vertex operator approach ${U_q(\hat{sl}(2))}$
Jimbo, Miki, Miwa, Nakayashiki \cite{Jimbo92,Nakayashiki94}~(1992).

\item $-1<\Delta<1$, $T=0$, $h=0$: qKZ equation
Jimbo and Miwa \cite{Jimbo96} (1996).

\item
 Rederivation by the quantum inverse scattering method.
Generalization  to the  XXZ  model with a magnetic f\/ield, $ T=0$
Kitanine, Maillet, Terras~\cite{Kitanine00} (1998).

\item Integral formula for f\/inite temperature and f\/inite magnetic f\/ield
G\"ohmann, Kl\"umper, Seel~\cite{Goem04}~(2004).

\end{enumerate}

  Correlation function for successive elementary block
\begin{gather*}
 \rho_n(\{\e_j,\e_j'\})=\frac{\langle\psi|\prod\limits_{j=1}^nE_j^{\e_j',\e_j}|\psi\rangle}{\langle\psi|\psi\rangle}\!, \\
 E^{+,+}_j=\left(\begin{matrix}1&0\\ 0&0\end{matrix}\right)_{[j]},\qquad E^{+,-}_j=\left(\begin{matrix}0&1\\ 0&0\end{matrix}\right)_{[j]}\!,\qquad
E^{-,+}_j=\left(\begin{matrix}0&0\\ 1&0\end{matrix}\right)_{[j]}\!,\qquad E^{-,-}_j=\left(\begin{matrix}0&0\\ 0&1\end{matrix}\right)_{[j]}\!,
\end{gather*}
is represented by $n$-fold integral.
For example the emptiness formation probability (EFP) for XXX chain at $T=h=0$ is
\begin{gather*}
P(n) \equiv \left\langle \left(S^z_1+\frac{1}{2} \right)  \left(S^z_2+\frac{1}{2}\right) \cdots
\left(S^z_n+\frac{1}{2} \right)  \right\rangle=\rho^{\ ++\cdots +}_{n ++\cdots +}  \nonumber  \\
\phantom{P(n)}{} = \left(-\pi\right)^{\frac{n(n-1)}{2}} 2^{-n} {\int_{-\infty}^{\infty}
{\rm d}^n \lambda }
\prod_{a>b}^n \frac{\sinh \pi(\lambda_a - \lambda_b)}{{\lambda_a-\lambda_b -  {\rm i}}}
\prod_{j=1}^{n}
\frac{{\left( \lambda_j -\frac{ {\rm i}}{2} \right)^{j-1}
\left(\lambda_j + \frac{{\rm i}}{2} \right)^{n-j}} }{\cosh^n \pi\lambda_j}.
\end{gather*}
Other arbitrary correlation functions  over  successive
$n$-sites have similar $n$-fold  integral representation.

\subsection[Boos-Korepin method to evaluate integrals]{Boos--Korepin method to evaluate integrals}

Here I introduce the details of direct factorization of multiple
integrals by Boos and Kore\-pin~\mbox{\cite{Boos01,Boos02}}.
\begin{enumerate}\itemsep=0pt
\item
Transform the integrand to a certain canonical form without changing
the integral value.

\item
Perform the integration using the residue theorem.

\end{enumerate}

Example:
\begin{gather*}
P(3)   \equiv   \prod_{j=1}^{3} \int_{-\infty -\frac{i}{2}}^{\infty-\frac{i}{2}}
\frac{{\rm d} \lambda_j}{2 \pi i} {U_{3}(\lambda_1,\lambda_2,\lambda_3)}
{T_{3}(\lambda_1,\lambda_2,\lambda_3)},
\\
{U_{3}(\lambda_1,\lambda_2,\lambda_3)}  \equiv
{\pi^6 \frac{\prod\limits_{1 \le k<j \le 3}
\sh \pi(\lambda_j-\lambda_k)}{\prod\limits_{j=1}^3 \sh^3 \pi \lambda_j}},  \nonumber \\
{T_{3}(\lambda_1,\lambda_2,\lambda_3)}  \equiv
{\frac{(\lambda_1+i)^2 \lambda_2 (\lambda_2+i) \lambda_3^2}
{(\lambda_2-\lambda_1-i)(\lambda_3-\lambda_1-i)(\lambda_3-\lambda_2-i)}}.
\end{gather*}
Details of transformation are given in Appendix~\ref{tran}.

In canonical form denominator is
\begin{gather*}
\prod_{k=1}^l(\lambda_{2k-1}-\lambda_{2k}),\qquad  l\le[n/2].
\end{gather*}
The $n$-dimensional integral is decomposed to one and two dimensional integrals.
Transform
${T_{3}(\lambda_1,\lambda_2,\lambda_3)}$ into a canonical form
${T_3^{\rm c}(\lambda_1,\lambda_2,\lambda_3)}$
\begin{gather*}
 T_{3}(\lambda_1,\lambda_2,\lambda_3) \sim T_3^{\rm c}(\lambda_1,\lambda_2,\lambda_3) =P_{0}^{(3)} +\frac{P_{1}^{(3)}}
{\lambda_2-\lambda_1} , \nonumber  \\
 P_0^{(3)} = - 2 \lambda_2 \lambda_3^2, \qquad P_1^{(3)} =
\frac{1}{3} - i \lambda_1-i \lambda_3 - 2 \lambda_1 \lambda_3.
\end{gather*}
Perform the integration
\begin{gather*}
J_0^{(3)}  =  \prod_{j=1}^{3} \int_{-\infty -\frac{i}{2}}^{\infty-\frac{i}{2}} \frac{{\rm d} \lambda_j}{2 \pi i}
{U_{3}(\lambda_1,\lambda_2,\lambda_3)} P_{0}^{(3)} = \frac{1}{4},  \nonumber  \\
J_1^{(3)}  =  \prod_{j=1}^{3} \int_{-\infty -\frac{i}{2}}^{\infty-\frac{i}{2}} \frac{{\rm d} \lambda_j}{2 \pi i}
{U_{3}(\lambda_1,\lambda_2,\lambda_3)} \frac{P_{1}^{(3)}}{\lambda_2-\lambda_1}
= -\ln 2+ \frac{3}{8} \zeta(3),  \nonumber \\
P(3)  =   J_0^{(3)} + J_1^{(3)} = \frac{1}{4} - \ln 2+ \frac{3}{8} \zeta(3).
\end{gather*}
Thus second neighbor correlator was rederived from the integral formula.

For $P(4)=\r^{\ ++++}_{4++++}$ the integrand is
\begin{gather*}
\frac{(\lambda_1+i)^3(\lambda_2+i)^2\lambda_2(\lambda_3+i)\lambda_3^2\lambda_4^3}{(\lambda_{43}-i)(\lambda_{42}-i)
(\lambda_{41}-i)(\lambda_{32}-i)(\lambda_{31}-i)(\lambda_{21}-i)}.
\end{gather*}
Here we put $\lambda_{ab}\equiv\lambda_a-\lambda_b$.
Canonical form is
$P_0^{(4)} + P_1^{(4)}/\l_{21} + P_2^{(4)}/( \l_{21}\l_{43})$,
\begin{gather*}
 P_0^{(4)} = -\frac{34}{5}\l_2\l_3^2\l_4^3, \\
 P_1^{(4)} =
\l_1^2 (30 \l_3^2\l_4^3 + 30 i\l_3\l_4^3 - 16 \l_4^3 +
18 \l_3\l_4^2 + 8 \l_4)   \\
 \phantom{P_1^{(4)} =}{} +\l_1 (30 i\l_3^2\l_4^3 + 30 \l_3\l_4^3 -
16 i\l_4^3 + 18 i\l_3\l_4^2 -
4 \l_4^2 + 4 i\l_4)  \nn\\
\phantom{P_1^{(4)} =}{} -20 \l_3^2\l_4^3 - 20 i\l_3\l_4^3 +
\frac{54}{ 5}\l_4^3 - \frac{42}{5}\l_3\l_4^2 -
\frac{43}{10}i\l_4, \\
 P_2^{(4)} = 2\l_1^2\l_3^2 + 4i\l_1\l_3^2 - \frac{3}{2}\l_3^2
 - \frac{3}{ 2}\l_1\l_3 - i\l_3 + \frac{1}{5},
\\
P(4)  = \frac{1}{5} - 2 {\ln 2} + \frac{173}{60} {\zeta(3)} - \frac{11}{6} \ln 2 \cdot {\zeta(3)}
- \frac{51}{80} {\zeta^2(3)}
- \frac{55}{24} {\zeta(5)} + \frac{85}{24} {\ln 2} \cdot {\zeta(5)}.
\end{gather*}
In 2003, we
calculated $\r^{+-+-}_{+-+-}$ by Boos--Korepin method and obtained {all} the correlation
functions on ${4}$ lattice sites~\cite{Sakai03}.  Especially, the third-neighbor correlator is
\begin{gather*}
\left\langle S_{j}^{z} S_{j+3}^{z} \right\rangle
  = \frac{1}{12} - 3 {\ln 2} + \frac{37}{6} {\zeta(3)} - \frac{14}{3} {\ln 2} \cdot {\zeta(3)}
 - \frac{3}{2} {\zeta(3)^2}  - \frac{125}{24} {\zeta(5)} + \frac{25}{3} {\ln 2} \cdot {\zeta(5)}    \\
\phantom{\left\langle S_{j}^{z} S_{j+3}^{z} \right\rangle}{}
= -0.05024862725 \dots.
\end{gather*}
The other correlation functions for $n=4$ are
\begin{gather*}
 \left\langle S_{j}^{x} S_{j+1}^{x} S_{j+2}^{z} S_{j+3}^{z} \right\rangle
 = \frac{1}{240}+\frac{1}{12} {\ln 2}-\frac{91}{240} {\zeta(3)}
+ \frac{1}{6} {\ln 2} \cdot {\zeta(3)}\\
\hphantom{\left\langle S_{j}^{x} S_{j+1}^{x} S_{j+2}^{z} S_{j+3}^{z} \right\rangle=}{}
+ \frac{3}{80} {\zeta(3)^2}  + \frac{35}{96} {\zeta(5)} - \frac{5}{24} {\ln 2}
\cdot {\zeta(5)}, \nonumber \\
 \left\langle S_j^{x} S_{j+1}^{z} S_{j+2}^{x} S_{j+3}^{z} \right\rangle
 = \frac{1}{240}-\frac{1}{6} {\ln 2}+\frac{77}{120} {\zeta(3)}
- \frac{5}{12} {\ln 2} \cdot {\zeta(3)}\\
\hphantom{\left\langle S_j^{x} S_{j+1}^{z} S_{j+2}^{x} S_{j+3}^{z} \right\rangle=}{}
-\frac{3}{20} {\zeta(3)^2} -\frac{65}{96} {\zeta(5)}
+ \frac{5}{6} {\ln 2} \cdot {\zeta(5)}, \nonumber  \\
 \left\langle S_j^{x} S_{j+1}^{z} S_{j+2}^{z} S_{j+3}^{x} \right\rangle
 = \frac{1}{240}- \frac{1}{4} {\ln 2}+\frac{169}{240} {\zeta(3)}
-\frac{5}{12} {\ln 2} \cdot {\zeta(3)}\\
\hphantom{\left\langle S_j^{x} S_{j+1}^{z} S_{j+2}^{z} S_{j+3}^{x} \right\rangle=}{}
-\frac{3}{20} {\zeta(3)^2} -\frac{65}{96} {\zeta(5)}
+\frac{5}{6} {\ln 2} \cdot {\zeta(5)}, \nonumber  \\
 \left\langle S_j^{z} S_{j+1}^{z} S_{j+2}^{z} S_{j+3}^{z} \right\rangle
 =  \left\langle S_{j}^{x} S_{j+1}^{x} S_{j+2}^{z} S_{j+3}^{z} \right\rangle
 + \left\langle S_j^{x} S_{j+1}^{z} S_{j+2}^{x} S_{j+3}^{z} \right\rangle
+ \left\langle S_j^{x} S_{j+1}^{z} S_{j+2}^{z} S_{j+3}^{x} \right\rangle.
\end{gather*}
From these results we can reproduce the twisted correlation function in~\eqref{twist}.

In a similar way, {${P(5)}$} was calculated after very tedious calculations~\cite{BKNS02}
\begin{gather*}
P(5) = \frac{1}{6} - \frac{10}{3} {\ln 2}  + \frac{281}{24} {\zeta(3)}
- \frac{45}{2} {\ln 2} \cdot {\zeta(3)}
- \frac{489}{16} {\zeta(3)^2}
  - \frac{6775}{192} {\zeta(5)}
+ \frac{1225}{6} {\ln 2} \cdot {\zeta(5)}\\
\hphantom{P(5) =}{}
- \frac{425}{64} {\zeta(3)} \cdot {\zeta(5)} - \frac{12125}{256} {\zeta(5)^2}
  + \frac{6223}{256} {\zeta(7)} - \frac{11515}{64} {\ln 2} \cdot {\zeta(7)}
+  \frac{42777}{512} {\zeta(3)} \cdot {\zeta(7)}.
\end{gather*}
But the direct integral of other correlations for f\/ive sites is almost impossible.

\subsection{Algebraic approach and qKZ relation}

Next problem is to calculate {${ \langle S_j^z S_{j+4}^z \rangle}$} for
XXX model.
In principle, it's possible to calculate other f\/ive-dimensional integrals by use of Boos--Korepin method. It, however, will take tremendous amount of time.

We propose a dif\/ferent method  (``algebraic approach'') and obtain analytical form of
{${\langle S_j^z S_{j+4}^z \rangle}$}. This is a
generalization of the method by Boos, Korepin, Smirnov (2003) for $P(6)$~\cite{Boos03}.
We consider the density matrix $\r$ of successive $n$ sites of inhomogeneous six vertex model
with dif\/ferent spectral
parameter $z_j$ for $j$-th site
\begin{gather*}
 \lim_{z_i \to 0} \r_{n,\e_1,\ldots,\e_n}^{\e'_1,\ldots,\e'_n} (z_1,z_2,\dots,z_n)  =
\langle E_{\e_1}^{\e'_1}\cdots E_{\e_n}^{\e'_n}   \rangle, \nonumber  \\
  (E_{\e}^{\e'})_{s,s'} = \d_{\e,s}\d_{\e',s'}, \qquad  \e,\e'=\pm 1.
\end{gather*}
For $n=1$,and $2$ we have
\begin{gather*}
\r_{1,+}^{+}(z_1)  =\r_{1,-}^{-}(z_1)= \frac{1}{2},\qquad \r_{1,+}^{-}(z_1) =\r_{1,-}^{+}(z_1)= 0,
\\
\r_{2,++}^{++}(z_1,z_2)  = \frac{1}{4} + \frac{1}{6} {\omega(z_1-z_2)}, \nonumber \\
\r_{2,+-}^{+-}(z_1,z_2)  =  - \frac{1}{6} {\omega(z_1-z_2)}, \qquad
\r_{2,+-}^{-+}(z_1,z_2) = \frac{1}{3} {\omega(z_1-z_2)},
\end{gather*}
with
\begin{gather*}
\omega(x)   \equiv \frac{1}{2}+2 \sum_{k=1}^{\infty} (-1)^k k \frac{1-x^2}{k^2-x^2}
  =  \frac{1}{2}-2\big(1-x^2\big) \sum_{k=0}^{\infty}   x^{2 k} \zeta_a(2k+1),  \\
 \zeta_a(x)\equiv \sum_{n=1}^\infty (-1)^{n-1}n^{-x}=\big(1-2^{1-x}\big)\zeta(x), \qquad \zeta_a(1)=\ln 2, \\
 \omega(x+1)=-\frac{x(x+2)}{x^2-1}\omega(x)-\frac{3}{2}\frac{1}{ 1-x^2}, \qquad \omega(-x)=\omega(x), \qquad \omega(\pm i\infty)=0.
\end{gather*}
The general element of density matrix must satisfy the following algebraic relations.
\begin{itemize}\itemsep=0pt
\item
Translational invariance
\begin{gather*}
\ds{\r_{n,\e_1,\ldots,\e_n}^{\e'_1,\ldots,\e'_n}(z_1+x,\ldots,z_n+x)=
\r_{n,\e_1,\ldots,\e_n}^{\e'_1,\ldots,\e'_n}(z_1,\ldots,z_n)}.
\end{gather*}
\item
Transposition, negating and reverse-order relations
\begin{gather*}
 \r_{n,\e_1,\ldots,\e_n}^{\e'_1,\ldots,\e'_n}(z_1,\ldots,z_n)=
\r_{n,\e'_1,\ldots,\e'_n}^{\e_1,\ldots,\e_n}(-z_1,\ldots,-z_n)  \\
\phantom{\r_{n,\e_1,\ldots,\e_n}^{\e'_1,\ldots,\e'_n}(z_1,\ldots,z_n)}{}
 =\r_{n,-\e_1,\ldots,-\e_n}^{-\e'_1,\ldots,-\e'_n}(z_1,\ldots,z_n)=
\r_{n,\e_n,\ldots,\e_1}^{\e'_n,\ldots,\e'_1}(-z_n,\ldots,-z_1) .
\end{gather*}
\item
Intertwining relation
\begin{gather*}
R_{\tilde\e'_j\tilde\e'_{j+1}}^{\e'_j\e'_{j+1}}(z_j-z_{j+1})
\r_{\ldots\e_{j+1},\e_j\ldots}^{\ldots\tilde\e'_{j+1},
\tilde\e'_j\ldots}(\ldots z_{j+1},z_j\ldots) \\
\qquad{} =\r_{\ldots\tilde\e_j,\tilde\e_{j+1}\ldots}^
{\ldots\e'_j,\e'_{j+1}\ldots}(\ldots z_j,z_{j+1}\ldots)
R^{\tilde\e_j\tilde\e_{j+1}}_{\e_j\e_{j+1}}(z_j-z_{j+1}),  \\
\\
 R^{++}_{++}(z)=R^{--}_{--}(z)=1,\qquad  R^{+-}_{+-}(z)=R^{-+}_{-+}(z)=\frac{z}{z+1},\\
  R^{+-}_{-+}(z)=R^{-+}_{+-}(z)=\frac{1}{z+1}.
\end{gather*}
\item
Reduction relation
\begin{gather*}
\r_{n,+,\e_2,\ldots,\e_n}^{+,\e'_2,\ldots,\e'_n}(z_1,z_2,\ldots,z_n) +
\r_{n,-,\e_2,\ldots,\e_n}^{-,\e'_2,\ldots,\e'_n}(z_1,z_2,\ldots,z_n)
=\r_{n-1,\e_2,\ldots,\e_n}^{\e'_2,\ldots,\e'_n}(z_2,\ldots,z_n).
\end{gather*}
\item
First recurrent relation
\begin{gather*}
  \r_{n,\e_1,\e_2,\ldots,\e_n}^{\e'_1,\e'_2,\ldots,\e'_n}
(z+1,z,z_3, \ldots,z_n)=-\d_{\e_1,-\e_2}\e'_1\e_2 \r_{n-1,-\e'_1,\e_3,\ldots,\e_n}^
{\e'_2,\e'_3,\ldots,\e'_n}(z,z_3,\ldots,z_n),  \\
  \r_{n,\e_1,\e_2,\ldots,\e_n}^{\e'_1,\e'_2,\ldots,\e'_n}
(z-1,z,z_3, \ldots,z_n) =-\d_{\e'_1,-\e'_2}\e_1\e'_2
\r_{n-1,\e_2,\e_3,\ldots,\e_n}^
{-\e_1,\e'_3,\ldots,\e'_n}(z,z_3,\ldots,z_n).
\end{gather*}
\item
  Second recurrent relation
\begin{gather*}
\lim_{z_1\rightarrow i\infty}
\r_{n,\e_1,\e_2,\ldots,\e_n}^{\e'_1,\e'_2,\ldots,\e'_n}
(z_1,z_2,\ldots,z_n)=
\delta_{\e_1,\e'_1} \frac{1}{2} \r_{n-1,\e_2,\ldots,\e_n}^
{\e'_2,\ldots,\e'_n}(z_2,\ldots,z_n).
\end{gather*}
\item
Identity relations
\begin{gather*}
 \sum_{\substack{\e_1,\ldots,\e_n\\ \sum_i \e'_i = \sum_i \e_i}}
\r_{n,\e_1,\ldots,\e_n}^{\e'_1,\ldots,\e'_n}(z_1,\ldots,z_n)
= \sum_{\substack{\e'_1,\ldots,\e'_n\\
\sum_i \e'_i = \sum_i \e_i}}
\r_{n,\e_1,\ldots,\e_n}^{\e'_1,\ldots,\e'_n}(z_1,\ldots,z_n) \nonumber \\
\qquad{} =\r_{n,+,\ldots,+}^{+,\ldots,+}(z_1,\ldots,z_n)
=\r_{n,-,\ldots,-}^{-,\ldots,-}(z_1,\ldots,z_n).
\end{gather*}
\end{itemize}

If we assume $\r_3$ as follows
\begin{gather*}
\r_{3,+++}^{\ +++}(z_1,z_2,z_3)  = \frac{1}{8} + {A(z_1,z_2|z_3)} {\omega(z_1-z_2)}
+ {A(z_1,z_3|z_2)} {\omega(z_1-z_3)}\\
\phantom{\r_{3,+++}^{\ +++}(z_1,z_2,z_3)  =}{} + {A(z_2,z_3|z_2)} \omega(z_2-z_3),   \nonumber \\
\r_{3,-++}^{\ -++}(z_1,z_2,z_3)  =\r_{2,++}^{\ ++}(z_2,z_3) - \r_{3,+++}^{\ +++}(z_1,z_2,z_3),   \nonumber \\
\r_{3,-+-}^{\ -+-}(z_1,z_2,z_3)  =\r_{2,-+}^{\ -+}(z_1,z_2) - \r_{3,-++}^{\ -++}(z_1,z_2,z_3), \quad \dots, \nonumber \\
{A(z_1,z_2|z_3)}  = \frac{(z_1-z_3)(z_2-z_3) -1}{12 (z_1-z_3)(z_2-z_3)},
\end{gather*}
these relations are satisf\/ied. In the homogeneous limit $z_j\to 0$ this gives the correct
correlation functions of XXX model.  In the homogeneous limit each term diverges but we have f\/inite
limiting number.  Then we can calculate the correlation functions of  arbitrary element of density matrix
using these algebraic relations, although the calculation become complicated.
We have calculated all the inhomogeneous correlation functions up to  $n \le 4$
from the multiple integrals and conf\/irmed these relations are fulf\/illed.

Further we have found the inhomogeneous correlation functions can be represented in terms of
$\omega$-function
\begin{gather*}
 \r_{n,\e_1,\ldots,\e_n}^{\ \e'_1,\ldots,\e'_n}(z_1,\ldots,z_n)=
\(\prod^n_{j=1}\frac{\delta_{\e_{j},\e'_{j}}}2\)+\sum_{m=1}^{\left[\frac{n}{ 2} \right]} \
\sum_{ 1 \le k_1<k_3<k_5<\cdots <k_{2m-1}<n, \  k_{2m}>k_{2m-1}} \\
\qquad {} {A_{\e_1,\ldots,\e_n}^{\e'_1,\ldots,\e'_n}( k_1,\dots ,k _{2m}|z_1,\ldots,z_n)}
 {\omega(z_{k_1}-z_{k_2}) \cdots \omega(z_{k_{2m-1}}-z_{k_{2m}})},   \\
\qquad{}  {A_{\e_1,\ldots,\e_n}^{\e'_1,\ldots,\e'_n}( k_1,\dots ,k _{2m}|z_1,\ldots,z_n)}= \frac{Q_{\e_1,\ldots,\e_n}^{\e'_1,\ldots,\e'_n}( k_1,\dots ,k _{2m}|z_1,\ldots, z_n )}{\prod^{'}_{i<j}(z_i-z_j)}  \\
 \qquad {} : \text{rational function of  ${z_1,\ldots,z_n}$}.
\end{gather*}
Denominator is
\begin{gather*}
\prod_{1\le j<k\le m} \!\! (z_{k_{2j-1}}-z_{k_{2l-1}})(z_{k_{2j-1}}-z_{k_{2l}})(z_{k_{2j}}-z_{k_{2l-1}})(z_{k_{2j}}-z_{k_{2l}})\prod_{l=1}^{2m}
\!\Bigg(\prod_{i\ne k_1,k_2,\dots,k_{2m}}\!\!(z_l-z_i)\Bigg).
\end{gather*}
The total exponent for this is $4nm-2m^2-2m$. The largest exponent for $z_i$ is $n-2$ for $i\in\{k_1,\dots,k_{2m}\}$ and $2m$ for $i\ne k_j$.
Numerator is also polynomials of $z_1,\dots,z_n$ which satisf\/ies the same exponent conditions. Unknowns are the coef\/f\/icients of each terms.
Number increases drastically as $n$, $m$ increases. Algebraic relations give the over complete linear equations.
By using mathematica we have unique solution of these equations.
By use of algebraic relations, we have calculated all the polynomials for  $n=5$
\begin{gather*}
 Q_{\e_1,\ldots,\e_5}^{\e'_1,\ldots,\e'_5}( k_1,k _{2}|z_1,\ldots, z_5 ) ,
\qquad Q_{\e_1,\ldots,\e_5}^{\e'_1,\ldots,\e'_5}( k_1,\ldots,k _{4}|z_1,\ldots, z_5 ).
\end{gather*}
By the memory problem this calculation stopped at $n=6$.

 Fourth-neighbor correlation function~\cite{Boos05}
\begin{gather*}
\left\langle S_{j}^{z} S_{j+4}^{z} \right\rangle
 = \frac{1}{12} - \frac{16}{3} {\ln 2}  + \frac{145}{6} {\zeta(3)} - 54 {\ln 2} \cdot {\zeta(3)}
 - \frac{293}{4} {\zeta(3)^2} - \frac{875}{12} {\zeta(5)} + \frac{1450}{3} {\ln 2} \cdot {\zeta(5)}
 \nonumber  \\
 \phantom{\left\langle S_{j}^{z} S_{j+4}^{z} \right\rangle=}{}  - \frac{275}{16} {\zeta(3)} \cdot {\zeta(5)} - \frac{1875}{16} {\zeta(5)^2}
 + \frac{3185}{64} {\zeta(7)} - \frac{1715}{4} {\ln 2} \cdot {\zeta(7)} +  \frac{6615}{32}  {\zeta(3)}
 \cdot {\zeta(7)}  \nonumber \\
\phantom{\left\langle S_{j}^{z} S_{j+4}^{z} \right\rangle}{}
= 0.034652776982 \dots,
\\
\zcor{j}{j+5}=
\frac{1}{12}-\frac{25}{3}\za{1}+\frac{800}{9}\za{3}-\frac{1192}{3}\za{1}\za{3}-\frac{15368}{9}
\za{3}^2
-608\za{3}^3\\
\phantom{\zcor{j}{j+5}= }{}
-\frac{4228}{9}\za{5}
+\frac{64256}{9}\za{1}\za{5}
-\frac{976}{9}\za{3}\za{5}
+3648\za{1}\za{3}\za{5}\\
\phantom{\zcor{j}{j+5}= }{}
-\frac{3328}{3}\za{3}^2\za{5}-\frac{76640}{3}\za{5}^2
+\frac{66560}{3}\za{1}\za{5}^2+\frac{12640}{3}\za{3}\za{5}^2\\
\phantom{\zcor{j}{j+5}= }{}
+\frac{6400}{3}\za{5}^3
+\frac{9674}{9}\za{7} +56952\za{3}\za{7}-\frac{225848}{9}\za{1}\za{7}\\
\phantom{\zcor{j}{j+5}= }{}
-\frac{116480}{3}\za{1}\za{3}\za{7}
 -\frac{35392}{3}\za{3}^2\za{7}+7840\za{5}\za{7}\\
\phantom{\zcor{j}{j+5}= }{}
 -8960\za{3}\za{5}\za{7}  -\frac{66640}{3}\za{7}^2+31360\za{1}\za{7}^2
-686\za{9}\\
\phantom{\zcor{j}{j+5}= }{}
+18368\za{1}\za{9}-53312\za{3}\za{9}+35392\za{1}\za{3}\za{9} \\
\phantom{\zcor{j}{j+5}= }{}
+16128\za{3}^2\za{9}+38080\za{5}\za{9}
-53760\za{1}\za{5}\za{9} \\
\phantom{\zcor{j}{j+5}}{}
= -0.03089036664760932\dots.
\end{gather*}
Using  this algebraic method we can calculate all the element of density
sub-matrix of successive 6-sites.
Longer system is quite dif\/f\/icult because of the memory and computing time problem.
We can calculate 6-th neighbor and 7-th neighbor correlations using the
generation function method.
They are represented by long polynomials of $\zeta_a$'s.  Here we write only numerical results~\cite{Sato05}
\begin{gather*}
\zcor{j}{j+6}= 0.02444673832795890\dots,\qquad
\zcor{j}{j+7}=-0.0224982227633722\dots.
\end{gather*}

\subsection{Calculation by continuous dimensions}

In a series of papers Boos, Jimbo, Miwa, Smirnov and Takeyama formulated these
algebraic calculation by the trace of continuous dimension of auxiliary space~\cite{BJMST1, BJMST2,BJMST3,BJMST4}
\begin{gather*}
(\rho_n)_{\e_1,\dots,\e_n}^{\ov{\e}_1,\dots,\ov{\e}_n} =
\bra{\rm vac}|(E^{\ov{\e}_1}_{\e_1})_1\cdots (E^{\ov{\e}_n}_{\e_n})_n|{\rm vac}\ket,\\
h_n(\e_1,\dots,\e_n,\ov{\e}_n,\dots,\ov{\e}_1) =(-1)^n\Bigg(\prod_{j=1}^n\ov{\e}_j\Bigg)(\rho_n)_{\e_1,\dots,\e_n}^{-\ov{\e}_1,\dots,-\ov{\e}_n},\\
s_n =\prod_{j=1}^n\frac{1}{2} \bigl(|+\rangle_j|-\rangle_{\ov{j}}-|-\rangle_j|+\rangle_{\ov{j}}\bigr),\qquad
h_n =\exp(\Omega_n)   s_n.
\end{gather*}
Density sub-matrix in $2^n$ dimensional space is mapped to a vector in $2^{2n}$
dimensional space.  $\Omega_n$~is an operator in this space.  Monodoromy matrix is def\/ined as
follows:
\begin{gather*}
L^{(0)}_j(\l) =\left(\l+\frac{1}{2}\right)I  \s_j^0+\frac{1}{2}(H \s^z_j+2E \s^+_j +2F \s^-_j), \\
T_n(\l) =L^{(0)}_{\ov{1}}(\l-z_1-1)\cdots L^{(0)}_{\ov{n}}(\l-z_n-1)L^{(0)}_n(\l-z_n)\cdots L^{(0)}_1(\l-z_1).
\end{gather*}
Here $I$, $H$, $E$, $F$ are $d\times d$ matrices
\begin{gather}
I_{i,j}=\d_{i,j},\qquad H_{i,j}=(d+1-2i)\d_{i,j},\qquad E_{i,j}=(i-1)\d_{i,j+1},\qquad F_{i,j}=(d-i)\d_{i+1,j}. \label{HEF}
\end{gather}
These satisfy commutation relations $[H,E]=-2E$, $[H,F]=2F$, $[E,F]=-H$.
$\s_j^z$, $\s_j^+$, $\s_j^-$ are Pauli operators in $2n$ spin space $j=1,\dots,n$, $\ov{1},\dots,\ov{n}$.
Especially at $d=2$ we have
\begin{gather*}
I=\left(\begin{matrix}1 &  0\\ 0 &  1\end{matrix}\right),\qquad H=\left(\begin{matrix}1 &  0\\ 0 &  -1\end{matrix}\right),\qquad
E=\left(\begin{matrix}0& 0\\ 1& 0\end{matrix}\right),\qquad
F=\left(\begin{matrix}0 & 1\\ 0 & 0\end{matrix}\right),
\end{gather*}
and
\begin{gather*}
T_n(\l)=\left(
\begin{matrix}
A(\l) & B(\l)\\ C(\l) & D(\l)
\end{matrix}\right).
\end{gather*}
The trace of any monomial of $I$, $E$, $H$ and $ F$ is a polynomial of dimension $d$.
One can calculate from the def\/inition~\eqref{HEF}. For example,
\begin{gather*}
{\rm Tr}_d I=d,\qquad {\rm Tr}_d H H=(d^3-d)/3,\qquad {\rm Tr}_d H=0.
\end{gather*}
The dimension $d$ is replaced by $\m-\nu $. Following~\cite{BJMST4}, the operator $\Omega_n$ is given by
\begin{gather*}
\Omega_n =\frac{1}{2}
\oint\frac{d \m}{2\pi i}\frac{d \nu}{2\pi i}\frac{(\m-\nu)\omega(\m-\nu)}
{(1-(\m-\nu)^2)^2\prod\limits_{j=1}^n(\m-z_j)(1-(\m-z_j)^2)(\nu-z_j)(1-(\nu-z_j)^2)} \nonum \\
\phantom{\Omega_n =}{} \times{\rm Tr}_{\m-\nu}T_n\left(\frac{\m+\nu}{2}\right)
(A(\m)D(\nu)+D(\m)A(\nu)-B(\m)C(\nu)-C(\m)B(\nu)),
\end{gather*}
where the integration path should surround all $z_j$ counter-clockwise.
In the homogeneous limit $z_j\to 0$ $\Omega$ becomes
\begin{gather*}
\Omega_n =\frac{1}{2}\oint\frac{d \m}{2\pi i}\frac{d \nu}{2\pi i}\frac{(\m-\nu)\omega(\m-\nu)}
{(1-(\m-\nu)^2)^2\m^n(1-\m^2)^n\nu^n(1-\nu^2)^n}  \\
\phantom{\Omega_n =}{} \times{\rm Tr}_{\m-\nu}T_n\left(\frac{\m+\nu}{2}
\right)(A(\m)D(\nu)+D(\m)A(\nu)-B(\m)C(\nu)-C(\m)B(\nu)),
\end{gather*}
and the calculation becomes very simple.  By this formulation we could calculate all the elements of
density sub-matrix at $n=7$. Calculating the eigenvalues of matrix, we can calculate the
von Neumann entropy (entanglement entropy) up to seven sites,
\begin{gather*}
S(n)\equiv-{\rm tr}\rho_n\log_2\rho_n
=-\sum_{\alpha=1}^{2^n}\omega_{\alpha}\log_2\omega_{\alpha},
\end{gather*}
where $\omega_{\alpha}$ are eigenvalues of density sub-matrix $\rho_n$. In Table~\ref{tb:von} $S(n)$
is given up to $n=7$.

\begin{table}[t]
\centering
\caption{von Neumann entropy $S(n)$ of a f\/inite sub-chain of length $n$.}
\label{tb:von}
\begin{tabular}{@{\hspace{\tabcolsep}\extracolsep{\fill}}cccc}
\hline
$S$(1)&$S$(2)&$S$(3)&$S$(4)\\
\hline
1&1.3758573262887466&1.5824933209573855&1.7247050949099274\\
\hline
\end{tabular}

\bigskip

\begin{tabular}{@{\hspace{\tabcolsep}\extracolsep{\fill}}ccc}
\hline
$S$(5)&$S$(6)&$S$(7)\\
\hline
1.833704916848315&1.922358833819333&1.997129812895912\\
\hline
\end{tabular}
\end{table}

\subsection{Generalization to XXZ model}

In 2003, Kato, Shiroishi, Takahashi, Sakai have generalized the Boos--Korepin method
to the XXZ models with an anisotropy parameter {${|\Delta| \le 1}$} for successive
three sites~\cite{Kato03}. For example {${P(n)}$} is represented as follows:
\begin{gather*}
  P(n)= \left(- \nu \right)^{-\frac{n(n-1)}{2}} {\int_{-\infty}^{\infty}
\frac{{\rm d} x_1}{2 \pi} \cdots \int_{-\infty}^{\infty}
\frac{{\rm d} x_n}{2 \pi}} \ \  \prod_{a>b} \frac{\sinh (x_a - x_b)}
{\sinh \left( \left(x_a-x_b - {\rm i} \pi \right) \nu \right)}  \nonumber \\
 \phantom{P(n)=}{}  \times  \prod_{j=1}^{n} \frac{\sinh^{n-j} \left( \left(x_j +\frac{ {\rm i} \pi}{2} \right)  \nu \right)
\sinh^{j-1} \left( \left(x_j - \frac{{\rm i} \pi}{2} \right) \nu \right)}{\cosh^n x}.  \label{Jimbo-Miwa}
\end{gather*}
Here ${ \Delta = \cos (\pi \nu)}$.
Similar integral representations for any arbitrary correlation function for successive
three sites were calculated.

\begin{figure}[t]
\centering
\includegraphics{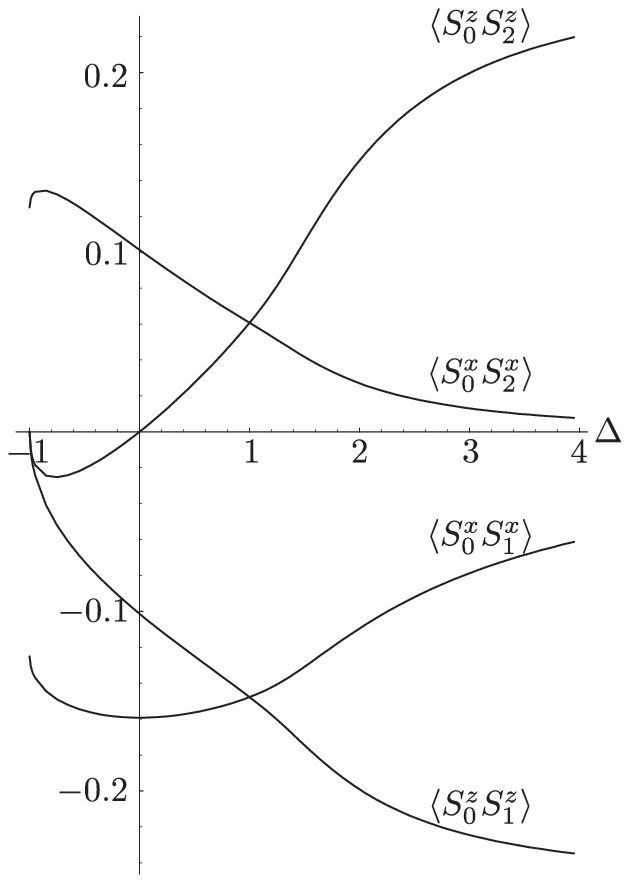}
\caption{The nearest-neighbor and the next nearest neighbor correlation functions for the XXZ chain.
We calculated $\bra S_j^zS_{j+1}^z\ket$, $\bra S_j^xS_{j+1}^x\ket$, $\bra S_j^zS_{j+2}^z\ket$
and $\bra S_j^xS_{j+2}^x\ket$.}\label{fig:fig1}
\end{figure}

\begin{figure}[t]
\centering
\includegraphics{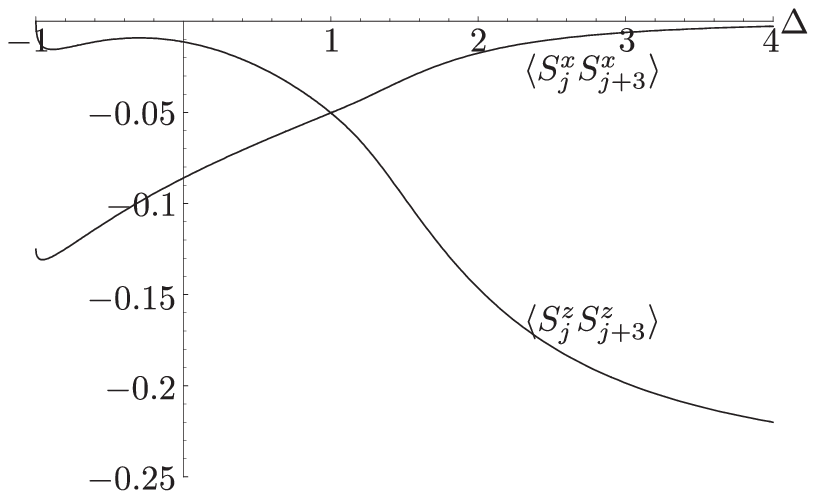}
\caption{The third-neighbor correlation functions for the XXZ chain.}\label{fig:fig2}
\end{figure}

Nearest-neighbor correlation functions
\begin{gather*}
\langle S_{j}^x S_{j+1}^x \rangle  = \frac{1}{4 \pi {s_1}} {\zeta_{\nu}(1)}
+\frac{{c_1}}{4 \pi^2} {\zeta'_{\nu}(1)}, \qquad
\langle S_{j}^z S_{j+1}^z \rangle = \frac{1}{4}- \frac{{c_1}}{2 \pi {s_1}} {\zeta_{\nu}(1)}
- \frac{1}{2 \pi^2} {\zeta'_{\nu}(1)}.
\end{gather*}
Next nearest-neighbor correlation functions
\begin{gather*}
\langle S_{j}^x S_{j+2}^x \rangle  = \frac{1}{2 \pi {s_2}} {\zeta_{\nu}(1)}+\frac{{c_2}}{4 \pi^2}
{\zeta'_{\nu}(1)}
- \frac{3(1-{c_2}){c_2}}{8 \pi {s_2}} {\zeta_{\nu}(3)}
-\frac{{s_1}^2}{8 \pi^2} {\zeta'_{\nu}(3)}, \nonumber  \\
\langle S_{j}^z S_{j+2}^z \rangle
= \frac{1}{4} - \frac{1+2 {c_2}}{\pi {s_2}} {\zeta_{\nu}(1)}
- \frac{1}{2 \pi^2} {\zeta'_{\nu}(1)}
+ \frac{3 {s_1}}{4 \pi {c_1}} {\zeta_{\nu}(3)}
+ \frac{1-{c_2}}{8 \pi^2} {\zeta'_{\nu}(3)}.
\end{gather*}
Here
\begin{gather}
{c_j}  := \cos \pi j \nu, \qquad s_j := \sin \pi j \nu,\nonumber
\\
{\zeta_{\nu}(j)}  := \int_{-\infty - \frac{\pi i}{2} }^{\infty - \frac{\pi i}{2}} {\rm d} x \frac{1}{{\rm sinh} x}
\frac{\cosh \nu x}{\sinh^j \nu x},\qquad
{\zeta'_{\nu}(j)}  := \int_{-\infty - \frac{\pi i}{2} }^{\infty - \frac{\pi i}{2}} {\rm d} x \frac{1}{{\rm sinh} x}
\frac{\partial }{\partial \nu} \frac{\cosh \nu x}{\sinh^j \nu x}.\label{zetas}
\end{gather}
Replacing {${\nu \to i \eta/\pi}$}, we can also get the correlation functions in the massive region  $\Delta = \cosh \eta >1$ \cite{Taka04}.
Third neighbor correlations is also expressed by functions
$\zeta_{\nu}$ and $\zeta'_{\nu}$,   although the expression becomes more complicated~\cite{Kato04}.
In Figs.~\ref{fig:fig1}  and~\ref{fig:fig2}, the nearest neighbor, the second neighbor and the third neighbor correlations are shown as functions of $\Delta$.

\section{Simplif\/ied thermodynamic Bethe ansatz equation}\label{sim}

Simplif\/ied thermodynamic Bethe ansatz equation at temperature $T$~\cite{mtaka01d,mtaka01e} is
\begin{gather}
u(x)-2\cosh(h/T)-\oint \frac{\theta}{2}\left[\coth\frac{\theta}{2}(x-y-2i)
\exp\left(-\frac{2\pi J\sin \theta }{\theta T}a_1(y+i)\right) \right.\nonumber\\
\left. \qquad {}+ \coth\frac{\theta}{2}(x-y+2i)\exp\left(-\frac{2\pi J\sin\theta }{\theta T}a_1(y-i)\right)\right]
\frac{1}{u(y)}\frac{d y}{2\pi i}=0,  \label{sTBA}
\end{gather}
where
\begin{gather*}
a_1(x)\equiv \frac{\theta\sin \theta}{2\pi (\cosh \theta x-\cos \theta)}.
\end{gather*}
Free energy per site is
\begin{gather*}
f=-T\ln u(0).
\end{gather*}

In this section we look for analytic solution for XY case $\Delta=0$, $\theta=\pi/2$. For this case equation \eqref{sTBA} becomes
\begin{gather*}
u(x)-2\cosh(h/T)-\oint \frac{\pi}{4}\tanh\frac{\pi}{4}(x-y) 2\cos\left(\frac{J}{T\sinh \pi y/2}\right)\frac{1}{u(y)}\frac{d y}{2\pi i}=0.
\end{gather*}
Putting $X=\tanh \pi x/4$, $Y=\tanh \pi y/4$ and using $u(-Y)=u(Y)$ we have
\begin{gather}
u(X)-2\cosh(h/T)+2(1-X^2)\!\oint \cos\left(\frac{J}{2 T}\frac{1\!-\!Y^2}{Y}\right)\!\frac{Y}{(1\!-\!X^2 Y^2)( 1\!-\!Y^2) u(Y)}\frac{d Y}{2\pi i}=0.\label{sXY}
\end{gather}
Consider the Fourier transform of following function
\begin{gather*}
g(\theta)=\frac{1}{2}\ln\left(4\left(\cosh^2\left( \frac{J}{2 T}\sin \theta\right)+\sinh^2(h/T)\right)\right) , \qquad
a_j\equiv \int_0^{2\pi}g(\theta)\cos(2 j\theta)\frac{d\theta}{ 2\pi}.
\end{gather*}
Assume that
\begin{gather}
\ln u(X)=a_0+2\sum_{j=1}^\infty a_j X^{2 j}.\label{logu}
\end{gather}
We can show that this satisf\/ies (\ref{sXY}). This series is convergent at $|X|\le 1$.
\begin{gather*}
 \ln u(e^{i\theta})+\ln u(e^{-i\theta})=2 a_0+2\sum_{j=1}^\infty a_j\cos(2j\theta) \\
 \qquad  {} =2g(\theta)=\ln\left(4\left(\cosh^2\left( \frac{J}{2T}\sin \theta\right)+\sinh^2(h/T)\right)\right).
\end{gather*}
Then we have
\begin{gather}
u(X)u(1/X)
=2\cos\left( \frac{J}{2 T}(X-1/X)\right)+2+4\sinh^2(h/T).\label{rel1}
\end{gather}
The function $u(X)$ has zeros at $X=\pm \beta_{j,\pm}$, where $\beta_{j,\pm}=\alpha_{j,\pm}+\sqrt{1+\alpha_{j,\pm}^2}$,  and $\alpha_{j,\pm}=
\frac{2\pi T}{J}(j-\frac{1}{2})\pm \frac{2 hi}{J}$. One can show that $|\beta_{j,\pm}|>1$.
$u(X)$ should not have zeros at $|X|<1$.
Then we have
$u(X)={\rm const}\cdot  \prod\limits_{j=1}^\infty (1-\frac{X^2}{\beta_{j, +}^2})(1-\frac{X^2}{ \beta_{j, -}^2})$. By the condition
$u(1)=2\cosh(h/T)$ we have inf\/inite product expansion of $u(X)$
\begin{gather}
u(X)=2\cosh(h/T)\prod_{j=1}^\infty \frac{\big(1-\frac{X^2}{\beta_{j, +}^2}\big)
\big(1-\frac{X^2}{\beta_{j, -}^2}\big)}{\big(1-\frac{1}{\beta_{j, +}^2}\big)\big(1-\frac{1}{\beta_{j, -}^2}\big)}.\label{u}
\end{gather}
From \eqref{rel1} we have
\begin{gather*}
2\cos\left(\frac{J}{2T}(Y-1/Y)\right)/u(Y)=u(1/Y)-\big(2+4\sinh^2(h/T)\big)/u(Y).
\end{gather*}
Then we can write $u(X)=2\cosh(h/T)+v(X^2)(1-X^2)$, $v(X)=\sum\limits_{j=0}^\infty d_j X^j$.  Convergence radius of $v(X)$ is inf\/inite
\begin{gather*}
 {\rm l.h.s. \ of \ (\ref{sXY})} = u(X)-2\cosh(h/T)+\big(1-X^2\big)\oint u(1/Y)\frac{Y}{(1-X^2Y^2)( 1-Y^2)}\frac{d Y}{\pi i} \\
 \hphantom{{\rm l.h.s. \ of \ (\ref{sXY})}}{}
 =\big(1-X^2\big)\left[v(X^2)-\oint v\big(1/Y^2\big)\frac{1}{Y(1-X^2Y^2)}\frac{d Y}{ 2\pi i}\right] \\
 \hphantom{{\rm l.h.s. \ of \ (\ref{sXY})}}{}
 =\big(1-X^2\big)\sum_{j=0}^\infty d_j\left(X^{2j}-\oint \frac{1}{Y^{2j+1}(1-X^2Y^2)}
\frac{d Y}{2\pi i}\right)=0.
\end{gather*}
Thus we have proved that \eqref{u}, \eqref{logu} satisf\/ies the equation \eqref{sXY}.
The free energy
\begin{gather*}
-T\ln u(0)=-T a_0
\end{gather*}
coincides with the known result~\cite{TakahashiBook}.

\section{Summary}

For $J<0$, $\Delta=1$, $T=h=0$ we obtained the factorized form of density sub-matrix
up to $n=7$. The entanglement entropy for seven sites is new result of this paper.  Up to
six sites we published in~\cite{Sato05}.

The six-th neighbor and the seven-th neighbor correlations are calculated by the generating function
method for $J<0$, $\Delta=1$, $T=h=0$~\cite{Sato05}.

For arbitrary $\Delta$, $T=h=0$ we obtained the factorized form up to $n=4$.
Correlations are given by two transcendental functions $\zeta_\nu(j)$ and  $\zeta_\nu'(j)$ with
$j=1,3,5,\dots$ def\/ined by~\eqref{zetas}.  For correlations of $n\ge 5$ the calculation becomes very tedious and no one has succeeded.

For simplif\/ied TBA equation, we obtained the analytic solution for XY limit $\Delta=0$.
Analytic solution for Ising limit was given in \cite{mtaka01d}.

\appendix
\section{Strong coupling expansion of the Hubbard model}\label{sce}
The Hubbard Hamiltonian is written as follows:
\begin{gather*}
{\cal H}=-t\sum_{<ij>}\sum_\sigma \big(c_{i\sigma}^\dagger c_{j \sigma}
+c_{j\sigma}^\dagger c_{i \sigma}\big)
+U\sum_{i=1}^{N_a} c_{i\uparrow}^\dagger c_{i\uparrow}
c_{i\downarrow}^\dagger c_{i\downarrow}.
\end{gather*}
If we treat the interaction term as main Hamiltonian and hopping therm as
perturbation in the half-f\/illed case
\begin{gather*}
 {\cal H}_0=U\sum n_{i\uparrow}n_{i\downarrow},\\
{\cal H}_1=-\sum_{\sigma}\sum_{i<j}t_{i,j}(c_{i\sigma}^\dagger c_{j\sigma}
+c_{j\sigma}^\dagger c_{i\sigma}),
\end{gather*}
the ef\/fective Hamiltonian becomes as follows:
\begin{gather*}{\cal H}_{\rm ef\/f}=\sum_{i<j}
\frac{t_{ij}t_{ji}}{U}(\sigma_i\cdot\sigma_j-1)
 +U^{-3}\Biggl[\sum_{i<j}t_{ij}^4(1- \sigma_i\cdot\sigma_j)+
\sum_{i<k}t_{ij}^2t_{jk}^2(\sigma_i\cdot\sigma_k-1)\\
\phantom{{\cal H}_{\rm ef\/f}=}{}
+\sum_{i<j<l,i<k,k\ne j,l}
t_{ij}t_{jk}t_{kl}t_{li}(5(\sigma_j\cdot\sigma_k)(\sigma_i\cdot\sigma_l)
+5(\sigma_i\cdot\sigma_j)(\sigma_k\cdot\sigma_l)
-5(\sigma_j\cdot\sigma_k)(\sigma_i\cdot\sigma_l) \\
\phantom{{\cal H}_{\rm ef\/f}=}{}
- \sigma_i\cdot\sigma_j- \sigma_j\cdot\sigma_k- \sigma_k\cdot\sigma_l
- \sigma_l\cdot\sigma_i- \sigma_i\cdot\sigma_k- \sigma_j\cdot\sigma_l+1)\Bigg], \\
 \sigma_i\cdot\sigma_j=4{\bf S}_i\cdot{\bf S}_j.
\end{gather*}
For one-dimensional half-f\/illed case the four spin term disappears and the ef\/fective Hamiltonian becomes
\begin{gather}
\frac{t^2}{U}\sum_i(4{\bf S}_i\cdot{\bf S}_{i+1}
-1)+\frac{t^4}{U^3}\sum_i\bigl\{4(1-4{\bf S}_i\cdot{\bf S}_{i+1})
+(4{\bf S}_i\cdot{\bf S}_{i+2}-1)\bigr\} +O\left(\frac{t^6}{U^5}\right).\label{effect}
\end{gather}
On the other hand exact ground state energy per site  is expanded as~\cite{liebwu68,Taka71b}
\begin{gather}
 e=-4|t|\int^\infty_0\frac{J_0(\omega)J_1(\omega)d\omega}{
\omega[1+\exp(2U'\omega)]}\nonumber\\
\phantom{e} =-4|t|\left[\left(\frac{1}{2}\right)^2\ln 2 \,  U^{'-1}
-\left(\frac{1\cdot 3}{2\cdot 4}\right)^2\frac{\zeta(3)}{3}\left(1-\frac{1}{2^2}\right)U^{'-3}+\cdots \right], \label{huben}\\
 U'\equiv U/(4|t|).\nonumber
\end{gather}
Comparing the f\/irst term of \eqref{effect} and \eqref{huben}, we get  nearest neighbor
correlation \eqref{nn38}. From the second term we get the second neighbor
correlation \eqref{nnn}.

\section[Transformation to canonical form in case of $P(3)$]{Transformation to canonical form in case of $\boldsymbol{P(3)}$}\label{tran}
\vspace{-4mm}

\begin{gather*}
T_3=\frac{(\lambda_1+i)^2(\lambda_2+i)\lambda_2\lambda_3^2}{(\lambda_3-\lambda_1-i)(\lambda_3-\lambda_2-i)
(\lambda_2-\lambda_1-i)}.
\end{gather*}
is decomposed to the following three terms,
\begin{gather*}
 =\frac{i(\lambda_1+i)^2(\lambda_2+i)\lambda_2\lambda_3^2}{(\lambda_3-\lambda_1-i)(\lambda_2-\lambda_1-i)}
+\frac{i(\lambda_1+i)^2(\lambda_2+i)\lambda_2\lambda_3^2}{(\lambda_3-\lambda_1-i)
(\lambda_3-\lambda_2-i)}
 -\frac{i(\lambda_1+i)^2(\lambda_2+i)\lambda_2\lambda_3^2}{(\lambda_3-\lambda_2-i)(\lambda_2-\lambda_1-i)}.
\end{gather*}
Using the antisymmetry of $U_3(\l_1,\l_2,\l_3)$ the f\/irst and the second terms are simplif\/ied as follows:
\begin{gather*}
{\rm the\ f\/irst\  term}
\sim\lambda_1^2\lambda_2-\ds\frac{(\lambda_1+i)^3\lambda_3}{\lambda_2-\lambda_1-i},
\\
 {\rm the\ second\  term}
\sim\lambda_1^2\lambda_2-\frac{(\lambda_1+i)^3(\lambda_3+i)}{\lambda_2-\lambda_1-i}.
\end{gather*}
The third term is transformed as follows
\begin{gather*}
 {\rm the\ third \  term}
\sim-\lambda_1^2\lambda_2
-i\frac{(\lambda_1+i)^3(\lambda_3+i)^2}{\lambda_2-\lambda_1-i}\!
+i\frac{(\lambda_1+i)^3\lambda_3^2}{\lambda_2-\lambda_1-i} \!
 -\frac{i(\lambda_1+i)^3\lambda_3^3}{(\lambda_3-\lambda_2-i)(\lambda_2-\lambda_1-i)}.
\end{gather*}
Then $T_3$ is transformed as follows:
\begin{gather*}
T_3\sim -\lambda_2\lambda_3^2
-\frac{i(\lambda_1+i)^3\lambda_3^3}{(\lambda_3
-\lambda_2-i)(\lambda_2-\lambda_1-i)}.
\end{gather*}
We should note that the pole at $\l_3=0$ and $\l_1=-i$ of $U_3$ is canceled by numerator of
the second term.  So we can change the integration path $\l_1\to\l_1-i$ and $\l_3\to\l_3+i$
\begin{gather*}
T_3\sim -\lambda_2\lambda_3^2
-\frac{i\lambda_1^3(\lambda_3+i)^3}{(\lambda_3-\lambda_2)(\lambda_2-\lambda_1)}.
\end{gather*}
Using the antisymmetry of $U$ we have
\begin{gather*}
\sim -\lambda_2\lambda_3^2-\frac{3\lambda_1^2\lambda_3^2+3i\lambda_1\lambda_3^2
+3i\lambda_1^2\lambda_3-\lambda_3^2-\lambda_3\lambda_1-\lambda_1^2}{\lambda_2-\lambda_1}.
\end{gather*}
Thus the denominator is drastically simplif\/ied.  Using the following relations
\begin{gather}
\frac{\lambda_1^2}{\lambda_2-\lambda_1}f(\lambda_3)\sim\left(\frac{-i\lambda_1
+\frac{1}{3}}{\lambda_2-\lambda_1}-
\frac{1}{3}(\lambda_1+i)\right)f(\lambda_3), \label{power}
\end{gather}
we can reduce the power of numerator,
\begin{gather*}
T_3\sim-2\lambda_2\lambda_3^2+\frac{\frac{1}{3}-i\lambda_1-i\lambda_3
-2\lambda_1\lambda_3}{\lambda_2-\lambda_1}.
\end{gather*}
Thus we have obtained the canonical form for $P(3)$.
The derivation of \eqref{power} is as follows:
\begin{gather*}
 \frac{\lambda_1^3}{\lambda_2-\lambda_1}f(\lambda_3)\sim -\frac{(\lambda_1+i)^3}{\lambda_2-\lambda_1-i}f(\lambda_3)
=\left(\frac{\lambda_2^3-(\l_1+i)^3}{\lambda_2-\lambda_1-i}-\frac{\lambda_2^3}{\lambda_2-\lambda_1-i}\right)f(\lambda_3)
 \\
\qquad{}
\sim \left(\frac{\lambda_2^3-(\l_1+i)^3}{\lambda_2-\lambda_1-i}+\frac{(\lambda_2+i)^3}{\lambda_2-\lambda_1}\right)f(\lambda_3)\sim \left(\frac{\lambda_2^3-(\l_1+i)^3}{\lambda_2-\lambda_1-i}+\frac{(\lambda_1+i)^3}{\lambda_2-\lambda_1}\right)f(\lambda_3).
\end{gather*}

\subsection*{Acknowledgements}
The author acknowledges to  A.~Kl\"umper, F.~G\"ohmann, H.~Boos, J.~Sato and M.~Shiroishi
for stimulating discussions. This work is f\/inancially supported by DFG.

\pdfbookmark[1]{References}{ref}
\LastPageEnding


\begin{thebibliography}{99}

\footnotesize\itemsep=0pt

\bibitem{Takahashi77}
Takahashi  M.,
Half-f\/illed Hubbard model at low temperature,
\href{http://dx.doi.org/10.1088/0022-3719/10/8/031}{{\it J.~Phys.~C}} {\bf 10} (1977), 1289--1301.

\bibitem{Yang69}
Yang C.N., Yang C.P.,
Thermodynamics of a one-dimensional system of bosons with repulsive delta-function interaction,
\href{http://dx.doi.org/10.1063/1.1664947}{{\it J.~Math. Phys.}} {\bf 10} (1969), 1115--1122.

\bibitem{Taka71}
Takahashi M.,
 One-dimensional Heisenberg model at f\/inite  temperature,
\href{http://dx.doi.org/10.1143/PTP.46.401}{{\it Prog. Theor. Phys.}} {\bf 46} (1971), 401--415.

\bibitem{Gaudin71}
Gaudin M.,
Thermodynamics of the Heisenberg--Ising ring for $\Delta\ge 1$,
\href{http://dx.doi.org/10.1103/PhysRevLett.26.1301}{{\it Phys. Rev. Lett.}}  {\bf 26} (1971),  1301--1304.

\bibitem{Taka72} Takahashi M., Suzuki M.,
One-dimensional anisotropic   Heisenberg model at f\/inite temperatures,
\href{http://dx.doi.org/10.1143/PTP.48.2187}{{\it Prog. Theor. Phys.}} {\bf 46} (1972), 2187--2209.

\bibitem{Koma89}
Koma T.,
Thermal Bethe-ansatz method for the spin-1/2 XXZ Heisenberg chain,
\href{http://dx.doi.org/10.1143/PTP.81.783}{{\it Prog. Theor. Phys.}} {\bf 81} (1989), 783--809.

\bibitem{Taka91}
Takahashi M.,
Correlation length and free energy of $S=1/2$ XXZ chain in magnetic f\/ield,
\href{http://dx.doi.org/10.1103/PhysRevB.44.12382}{{\it Phys. Rev.~B}}  {\bf 44} (1991), 12382--12394.

\bibitem{Klum93}
Kl\"umper A.,
Thermodynamics of the anisotropic spin-1/2 Heisenberg chain and related quantum chains,
\href{http://dx.doi.org/10.1007/BF01316831}{{\it Z.~Phys.~B}} {\bf 91} (1993), 507--519,
\href{http://arxiv.org/abs/cond-mat/9306019}{cond-mat/9306019}.

\bibitem{mtaka01d}
Takahashi M., Simplif\/ication of thermodynamic Bethe-ansatz equations,
in Physics and Combinatrix (Nagoya, 2000), World Sci. Publ., River Edge, NJ, 2001, 299--304,
\href{http://arxiv.org/abs/cond-mat/0010486}{cond-mat/0010486}.

\bibitem{mtaka01e}
Takahashi M.,  Shiroishi M.,  Kl\"{u}mper A.,
Equivalence of TBA and QTM,
\href{http://dx.doi.org/10.1088/0305-4470/34/13/105}{{\it J.~Phys.~A: Math. Gen.}}  {\bf 34} (2001), L187--L194,
\href{http://arxiv.org/abs/cond-mat/0102027}{cond-mat/0102027}.

\bibitem{shiro02}
Shiroishi M., Takahashi M.,
Integral equation generates high-temperature expansion of the Heisenberg chain,
\href{http://dx.doi.org/10.1103/PhysRevLett.89.117201}{{\it Phys. Rev. Lett.}} {\bf 89} (2002), 117201,  4~pages,
\href{http://arxiv.org/abs/cond-mat/0205180}{cond-mat/0205180}.

\bibitem{Tsuboi05}
Tsuboi Z., Takahashi M.,
Nonlinear integral equations for thermodynamics of the $U_{q}(\widehat{sl(r+1)})$ Perk--Schultz model,
\href{http://dx.doi.org/10.1143/JPSJ.74.898}{{\it J.~Phys. Soc. Japan}} {\bf 74} (2005), 898--904,
\href{http://arxiv.org/abs/cond-mat/0412698}{cond-mat/0412698}.

\bibitem{Hulthen38}
Hulth\'{e}n L.,
\"Uber das Austauschproblem eines Kristalles,
{\it  Ark. Mat. Astron. Fys.~A} {\bf 26} (1938), 1--105.

\bibitem{mura99}
Muramoto N., Takahashi M.,
Integrable magnetic model of two chains coupled by four-body interactions,
\href{http://dx.doi.org/10.1143/JPSJ.68.2098}{{\it J.~Phys. Soc. Japan}} {\bf 68} (1999),  2098--2104,
\href{http://arxiv.org/abs/cond-mat/9902007}{cond-mat/9902007}.

\bibitem{Jimbo92}
Jimbo M., Miki K., Miwa T., Nakayashiki A.,
Correlation functions of the XXZ model for $\Delta<-1$,
\href{http://dx.doi.org/10.1016/0375-9601(92)91128-E}{{\it Phys. Lett.~A}} {\bf 168} (1992), 256--263,
\href{http://arxiv.org/abs/hep-th/9205055}{hep-th/9205055}.

\bibitem{Nakayashiki94}
Nakayashiki A.,
Some integral formulas for the solutions of the $sl_2$ dKZ equation with level-4,
\href{http://dx.doi.org/10.1142/S0217751X94002326}{{\it Internat. J. Modern Phys.~A}} {\bf 9} (1994), 5673--5687.

\bibitem{Jimbo96}
Jimbo M., Miwa T.,
Quantum KZ equation with $|q| = 1$ and correlation functions of the XXZ model in the gapless regime,
\href{http://dx.doi.org/10.1088/0305-4470/29/12/005}{{\it J.~Phys.~A: Math. Gen.}} {\bf 29} (1996), 2923--2958,
\href{http://arxiv.org/abs/hep-th/9601135}{hep-th/9601135}.

\bibitem{Kitanine00}
Kitanine N., Maillet J.M., Terras V.,
Correlation functions of the XXZ Heisenberg spin-1/2 chain in a~magnetic f\/ield,
\href{http://dx.doi.org/10.1016/S0550-3213(99)00619-7}{{\it  Nuclear Phys.~B}} {\bf 567} (2000), 554--582,
\href{http://arxiv.org/abs/math-ph/9907019}{math-ph/9907019}.

\bibitem{Goem04}
G\"ohmann F., Kl\"umper A., Seel A.,
Integral representations for correlation functions of the XXZ chain at f\/inite temperature,
\href{http://dx.doi.org/10.1088/0305-4470/37/31/001}{{\it J.~Phys.~A: Math. Gen.}} {\bf 37} (2004), 7625--7651,
\href{http://arxiv.org/abs/hep-th/0405089}{hep-th/0405089}.

\bibitem{Boos01}
Boos H.E., Korepin V.E.,
Quantum spin chains and Riemann zeta function with odd arguments,
\href{http://dx.doi.org/10.1088/0305-4470/34/26/301}{{\it J.~Phys.~A: Math. Gen.}} {\bf 34} (2001), 5311--5316,
\href{http://arxiv.org/abs/hep-th/0104008}{hep-th/0104008}.

\bibitem{Boos02}
Boos H.E., Korepin V.E.,
 Evaluation of integrals representing correlations in XXX Heisenberg spin chain,
in MathPhys Odyssey (2001), {\it Prog. Math. Phys.}, Vol.~23, Birkh\"auser Boston, Boston, MA, 2002, 65--108,
\href{http://arxiv.org/abs/hep-th/0105144}{hep-th/0105144}.

\bibitem{Sakai03}
Sakai  K., Shiroishi M., Nishiyama Y., Takahashi M.,
Third-neighbor correlators of a one-dimensional spin-1/2 Heisenberg antiferromagnet,
\href{http://dx.doi.org/10.1103/PhysRevE.67.065101}{{\it Phys. Rev.~E}} {\bf 67} (2003), 065101(R), 4~pages,
\href{http://arxiv.org/abs/cond-mat/0302564}{cond-mat/0302564}.

\bibitem{BKNS02}
Boos H.E., Korepin V.E., Nishiyama Y.,  Shiroishi M.,
Quantum correlations and number theory,
\href{http://dx.doi.org/10.1088/0305-4470/35/20/305}{{\it J.~Phys.~A: Math. Gen.}} {\bf 35} (2002), 4443--4451,
\href{http://arxiv.org/abs/cond-mat/0202346}{cond-mat/0202346}.

\bibitem{Boos03}
Boos H.E., Korepin V.E., Smirnov F.A.,
Emptiness formation probability and quantum Knizhnik--Zamolodchikov equation,
\href{http://dx.doi.org/10.1016/S0550-3213(03)00153-6}{{\it  Nuclear Phys.~B}} {\bf 658} (2003), 417--439,
\href{http://arxiv.org/abs/hep-th/0209246}{hep-th/0209246}.

\bibitem{Boos05}
Boos H.E.,  Shiroishi M., Takahashi M.,
First principle approach to correlation functions of spin-1/2 Heisenberg chain: fourth-neighbor correlators,
\href{http://dx.doi.org/10.1016/j.nuclphysb.2005.01.041}{{\it  Nuclear Phys.~B}} {\bf 712} (2005), 573--599,
\href{http://arxiv.org/abs/hep-th/0410039}{hep-th/0410039}.

\bibitem{Sato05}
Sato J., Shiroishi M., Takahashi M.,
Correlation functions of the spin-1/2 anti-ferromagnetic Heisenberg chain: exact calculation via the generating function,
\href{http://dx.doi.org/10.1016/j.nuclphysb.2005.08.045}{{\it  Nuclear Phys.~B}} {\bf 729}  (2005), 441--466,
\href{http://arxiv.org/abs/hep-th/0507290}{hep-th/0507290}.

\bibitem{BJMST1}
Boos H.,  Jimbo M., Miwa T., Smirnov F., Takeyama  Y.,
A recursion formula for the correlation functions of an inhomogeneous XXX model,
\href{http://dx.doi.org/10.1090/S1061-0022-06-00894-6}{{\it St.~Petersburg Math.~J.}} {\bf 17} (2005), 85--117,
\href{http://arxiv.org/abs/hep-th/0405044}{hep-th/0405044}.

\bibitem{BJMST2}
Boos H., Jimbo M., Miwa T., Smirnov F., Takeyama Y.,
Reduced qKZ equation and correlation functions of the XXZ model,
\href{http://dx.doi.org/10.1007/s00220-005-1430-6}{{\it Comm. Math. Phys.}}  {\bf 261} (2006), 245--276,
\href{http://arxiv.org/abs/hep-th/0412191}{hep-th/0412191}.

\bibitem{BJMST3}
Boos H., Jimbo M., Miwa T., Smirnov F., Takeyama Y.,
Traces on the Sklyanin algebra and correlation functions of the eight-vertex model,
\href{http://dx.doi.org/10.1088/0305-4470/38/35/003}{{\it J.~Phys.~A: Math. Gen.}} {\bf 38} (2005), 7629--7659,
\href{http://arxiv.org/abs/hep-th/0504072}{hep-th/0504072}.

\bibitem{BJMST4}
Boos H.,  Jimbo M.,  Miwa T.,  Smirnov F., Takeyama Y.,
Density matrix of a f\/inite sub-chain of the Heisenberg anti-ferromagnet,
\href{http://dx.doi.org/10.1007/s11005-006-0054-x}{{\it Lett. Math. Phys.}} {\bf 75} (2006), 201--208,
\href{http://arxiv.org/abs/hep-th/0506171}{hep-th/0506171}.

\bibitem{Kato03}
Kato G.,  Shiroishi M.,  Takahashi M., Sakai K.,
Next nearest-neighbor correlation functions of the spin-1/2 XXZ chain at critical region,
\href{http://dx.doi.org/10.1088/0305-4470/36/23/102}{{\it J.~Phys.~A: Math. Gen.}} {\bf 36} (2003), L337--L344,
\href{http://arxiv.org/abs/cond-mat/0304475}{cond-mat/0304475}.

\bibitem{Taka04}
Takahashi M.,  Kato G.,  Shiroishi M.,
Next nearest-neighbor correlation functions of the spin-1/2 XXZ chain at massive region,
\href{http://dx.doi.org/10.1143/JPSJ.73.245}{{\it J. Phys. Soc. Japan}} {\bf 73}  (2004), 245--253,
\href{http://arxiv.org/abs/cond-mat/0308589}{cond-mat/0308589}.

\bibitem{Kato04}
Kato G.,  Shiroishi M.,  Takahashi M.  Sakai K.,
Third-neighbour and other four-point correlation functions of spin-1/2 XXZ chain,
\href{http://dx.doi.org/10.1088/0305-4470/37/19/001}{{\it J.~Phys.~A: Math. Gen.}} {\bf 37}   (2004), 5097--5123,
\href{http://arxiv.org/abs/cond-mat/0402625}{cond-mat/0402625}.

\bibitem{TakahashiBook}
Takahashi M.,
Thermodynamics of one-dimensional solvable models,
Cambridge University Press, Cambridge, 1999.

\bibitem{liebwu68}
Lieb E.H., Wu F.Y.,
Absence of mott transition in an exact solution of the short-range, one-band model in one dimension,
\href{http://dx.doi.org/10.1103/PhysRevLett.20.1445}{{\it Phys. Rev. Lett.}} {\bf 20} (1968), 1445--1448, Erratum, \href{http://dx.doi.org/10.1103/PhysRevLett.21.192.2}{{\it Phys. Rev. Lett.}} {\bf 21}  (1968), 192.

\bibitem{Taka71b}
Takahashi M.,
On the exact ground state energy of Lieb  and Wu,
\href{http://dx.doi.org/10.1143/PTP.45.756}{{\it Prog. Theor. Phys.}} {\bf 45}  (1971), 756--760.

\end{thebibliography}
\end{document}